\documentclass[onecolumn,trackchanges]{aastex631}
\usepackage{amsmath}
\shorttitle{Synchrotron polarization of a hybrid distribution of relativistic thermal and nonthermal electrons}
\shortauthors{Cheng et al.}
\graphicspath{{./}{figures/}}
\begin{document}

\title{Synchrotron polarization of a hybrid distribution of relativistic thermal and nonthermal electrons in GRB prompt emission}

\correspondingauthor{Jirong Mao}
\email{jirongmao@mail.ynao.ac.cn}

\author[0009-0004-3324-8421]{Kangfa Cheng}
\affiliation{School of Mathematics and Physics, Guangxi Minzu University \\
Nanning 530006, People’s Republic of China}

\author{Jirong Mao}
\affiliation{Yunnan Observatories, Chinese Academy of Sciences \\ 
Kunming, 650011, People’s Republic of China}
\affiliation{Key Laboratory for the Structure and Evolution of Celestial Objects, Chinese Academy of Science \\
Kunming, 650011, People’s Republic of China }
\affiliation{Center for Astronomical Mega-Science, Chinese Academy of Science \\
20A Datun Road, Chaoyang District, Beijing 100012, People’s Republic of China}

\author{Xiaohong Zhao}
\affiliation{Yunnan Observatories, Chinese Academy of Sciences \\ 
Kunming, 650011, People’s Republic of China}
\affiliation{Key Laboratory for the Structure and Evolution of Celestial Objects, Chinese Academy of Science \\
Kunming, 650011, People’s Republic of China }
\affiliation{Center for Astronomical Mega-Science, Chinese Academy of Science \\
20A Datun Road, Chaoyang District, Beijing 100012, People’s Republic of China}

\author{Hongbang Liu}
\affiliation{School of Physical Science and Technology, Guangxi University \\
Nanning 530004, People’s Republic of China}

\author{Merlin Kole}
\affiliation{Space Science Center, University of New Hampshire \\
Durham, NH 03824, USA
} 

\author{Nicolas Produit}
\affiliation{Astronomy Department, University of Geneva \\
Versoix, Switzerland 
}

\author{Zhifu Chen}
\affiliation{School of Mathematics and Physics, Guangxi Minzu University \\
Nanning 530006, People’s Republic of China}




\begin{abstract}
Synchrotron polarization of relativistic nonthermal electrons in gamma-ray bursts (GRBs) has been widely studied. However, recent numerical simulations of relativistic shocks and magnetic reconnection have found that a more realistic electron distribution consists of a power-law component plus a thermal component, which requires observational validation. In this paper, we investigate synchrotron polarization using a hybrid energy distribution of relativistic thermal and nonthermal electrons within a globally toroidal magnetic field in GRB prompt emission. Our results show that, compared to the case of solely non-thermal electrons, the synchrotron polarization degrees (PDs) in these hybrid electrons can vary widely depending on different parameters and that the PD decreases progressively with frequency in the $\gamma$-ray, X-ray, and optical bands. The time-averaged PD spectrum displays a significant bump in the $\gamma$-ray and X-ray bands with the PDs higher than $\sim60\%$ if the thermal peak energy of electrons is much smaller than the conjunctive energy of electrons between the thermal and non-thermal distribution. The high synchrotron PD ($\gtrsim 60\%$) in the $\gamma$-ray and X-ray bands, which generally can not be produced by solely non-thermal electrons with typical power-law slopes, can be achieved by the hybrid electrons and primarily originates from the exponential decay part of the thermal component. Moreover, this model can roughly explain the PDs and spectral properties of some GRBs, where GRB 110301A with a high PD ($70_{-22}^{+22} \%$) may be potential evidence for the existence of relativistic thermal electrons.
\end{abstract}



\section{Introduction} \label{sec:intro}
Observed spectra of gamma-ray bursts (GRBs) are generally characterized by non-thermal broken power-law spectra (Band spectrum, \citealt{1993ApJ...413..281B}), whose radiation mechanism still remains a mystery.
The synchrotron radiation from power-law relativistic electrons is one of the most competitive mechanisms. The relativistic electrons can originate from internal shocks (e.g. \citealt{Rees+Meszaros+1994, Paczynski+Xu+1994}) or the dissipation of magnetic field (MF) energy (e.g. \citealt{Usov+1992, Thompson+1994, Spruit+etal+2001, Vlahakis+2003, Zhang+Yan+2011}).   
The power-law distribution of electrons is generally believed to be produced by the first-order or second-order Fermi acceleration \citep{Schlickeiser+1985, Stawarz+Petrosian+2008}. Such an electron distribution in a decaying MF model ($B \propto R^{-a}$) can successfully explain the observed GRB Band spectrum \citep{Uhm+Zhang+2014, Zhao+etal+2014}.
However, some recent numerical simulation results have shown that in addition to relativistic nonthermal electrons, relativistic thermal electrons can also be produced in some synchrotron sources. The energy distribution of relativistic thermal electrons 
has a Maxwellian form, which is usually produced by thermalization or isotropization of particle momenta. Some particle-in-cell (PIC) simulations of relativistic shocks have found that only a small fraction ($\sim 10\%$) of energy dissipated into the power-law distribution of electrons downstream of the relativistic shocks, and the majority of the energy is deposited into a Maxwellian distribution of electrons \citep{Spitkovsky+2008a, Spitkovsky+2008b, Martins+etal+2009, Sironi+Spitkovsky+2009}. \cite{Giannios+Spitkovsky+2009} calculated the synchrotron spectrum by considering such a hybrid energy distribution of the relativistic thermal and nonthermal electrons, whose results have shown that the model could explain some observed GRB spectra.
These results are based on the baryon-dominated relativistic shocks model. 
Similarly, the recent PIC simulations of magnetic field reconnection in the Poynting-flux-dominated outflow assume an initial Maxwellian distribution of electrons and found that the distribution evolves gradually from the initial thermal distribution to a combination of a thermal distribution at lower energies and a power-law distribution at higher energies \citep{Guo+etal+2014, Guo+etal+2015, Sironi+Spitkovsky+2014}. Thus the electron distribution with a thermal component plus a power-law component is a general form applicable to both the baryon-dominated and the Poynting-flux-dominated outflow.

Whether the relativistic electrons of synchrotron radiation are purely nonthermal electrons or hybrid electrons remains to be further studied. Polarization can serve as a probe of the two components. The synchrotron polarization of the relativistic nonthermal electrons has been widely studied \citep{Granot+konigl+2003, Nakar+etal+2003, Toma+etal+2009, Cheng+etal+2020, Cheng+etal+2024, Lan+etal+2019, Lan+etal+2020, Lan+etal+2021(2), Lan+etal+2021, Gill+etal+2020, Gill+etal+2021, Gill+Granot+2021, Gill+Granot+2024}. The time-averaged polarization degrees (PDs) of synchrotron emission from relativistic nonthermal electrons are from $\sim10\%$ to $\sim 60\%$ for various jet structures and MF models \citep{Cheng+etal+2020, Gill+etal+2021, Sui+Lan+2024}. 
The synchrotron polarization of the hybrid 
distribution of relativistic thermal and nonthermal electrons has been studied by \cite{Mao+Wang+2018} (hereafter "MW18") and \cite{Mao+etal+2018}. MW18 calculated the synchrotron PDs of the hybrid distribution of relativistic thermal and nonthermal electrons and found that the PDs are distributed over a very wide range. However, the PD calculated in MW18 is the intrinsic polarization and does not take any geometric effects into account. Considering the jet geometric effects and different MF models will lead to a significant change in the PDs, the results in MW18 will not be appropriate to describe the observed data directly. To utilize MW18 for the explanation of GRB polarization,
we calculate the synchrotron polarization of the hybrid 
electrons by considering the jet geometric effects in this paper. We consider a homogeneous top-hat jet with a large-scale toroidal MF model. This paper is organized as follows. Section \ref{sec:dis-ele} describes the mixture energy distribution of thermal and nonthermal electrons. Section \ref{sec:cal-pol} shows the models and calculations of synchrotron polarization, including energy-resolved polarization, instantaneous polarization, and time-averaged polarization. The summary and discussions are 
shown in Section \ref{sec: conclusion}.

\section{A hybrid energy distribution of relativistic thermal and nonthermal electrons} \label{sec:dis-ele}
\begin{figure}[ht!]
\plotone{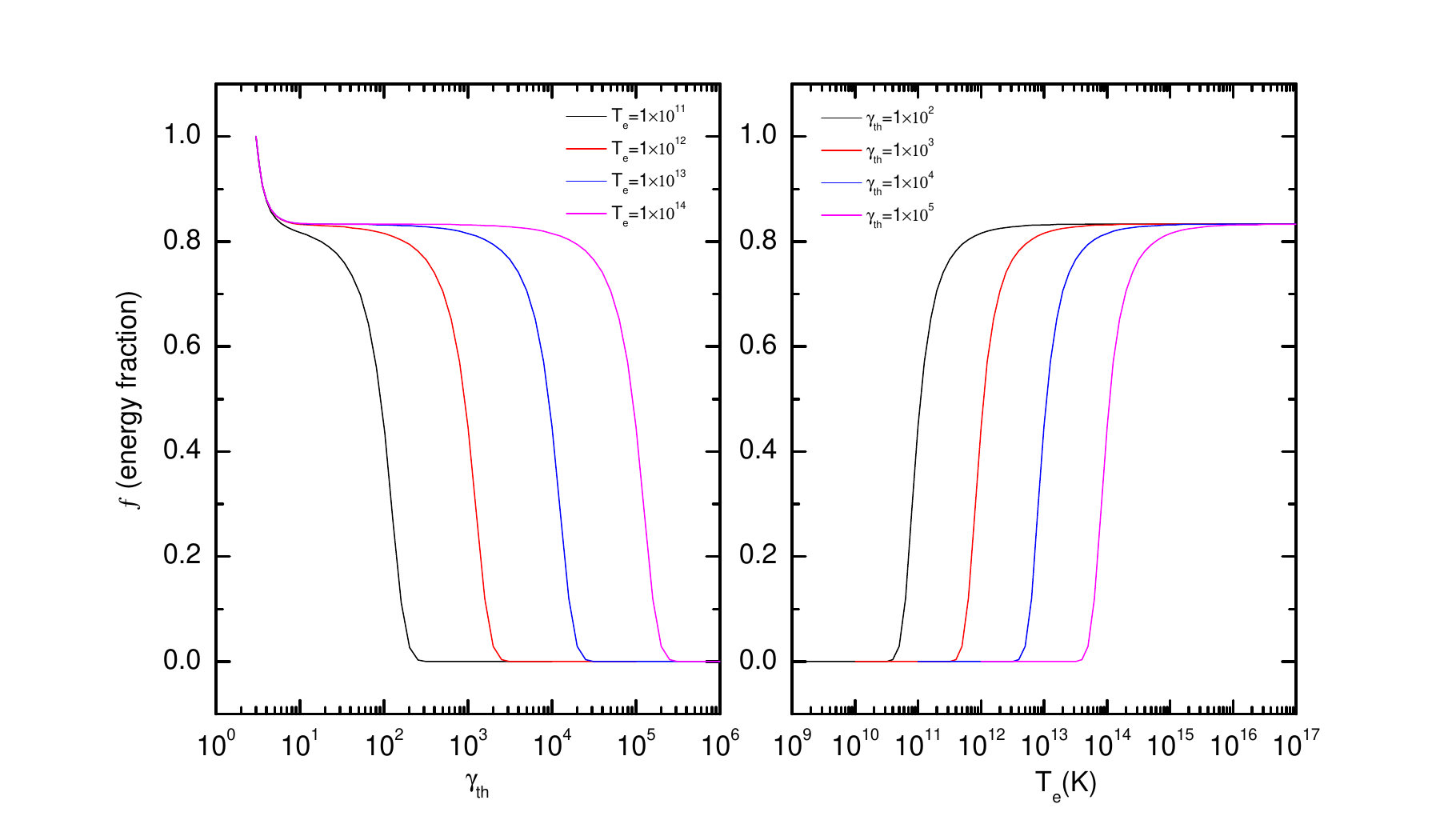}
\caption{Left panel: The energy fraction of nonthermal electrons with different conjunctive LFs ($\gamma_{th}$). Right panel: The energy fraction of nonthermal electrons with different temperatures of thermal electrons ($T_e$). \label{fig: frac}}
\end{figure}

Both PIC simulations of baryon-dominated relativistic shocks and Poynting-flux-dominated outflow have revealed that the resulting distribution of electrons is a combination of a thermal component and a non-thermal component. Here we use such an electron distribution in GRB prompt emission. Electrons with Lorentz factors (LFs) below a conjunctive LF ($\gamma_{th}$) are distributed as a Maxwellian form (thermal component), while the electrons with LFs higher than $\gamma_{th}$ are distributed as a power-law form (nonthermal component). The electron distribution can be written as \citep{Giannios+Spitkovsky+2009, Mao+Wang+2018}

\begin{equation} \label{ele-dis}
N_{e} (\gamma) =
\begin{cases}
N_{0} \gamma^{2} \exp(-\gamma/\Theta)/2\Theta^{3}   & (\rm{for \quad \gamma \leq \gamma_{th}}) \\
N_{0} \gamma_{th}^{2} \exp(-\gamma_{th}/\Theta)(\gamma/\gamma_{th})^{-p}/2\Theta^{3}  & (\rm{for \quad \gamma >\gamma_{th}})
\end{cases}
\end{equation} 

where $N_{0}$ is a constant, $\gamma$ is the LF of the electron, $\Theta= kT_{e}/m_{e}c^{2}$ is the characteristic temperature, and $p$ is the power-law index of the nonthermal electron energy distribution. We take $p=2.8$ in this paper. $T_{e}$ is the temperature of the relativistic thermal electrons, $k$ is the Boltzmann constant, $m_e$ is the electron mass, and $c$ is the speed of light. To better investigate the relations between the fraction of the thermal/nonthermal electron energy and the electron distribution, we define a parameter $f$ to describe the energy fraction of nonthermal electrons. It can be calculated as \citep{Giannios+Spitkovsky+2009, Mao+Wang+2018}

\begin{equation}
f=\frac{\int_{\gamma_{th}}^{\infty} \gamma N_{e}(\gamma,\Theta) d\gamma}{\int_{\gamma_{min}}^{\infty}\gamma N_{e}(\gamma,\Theta)d\gamma}.
\end{equation} \label{frac}

where $\gamma_{\rm{min}}$ is the minimum LF of the relativistic electrons and we take $\gamma_{\rm{min}}=3$ in this paper. According to Equation \ref{frac}, the nonthermal energy fraction $f$ is related to 
$\gamma_{th}$ and $T_e$. We present the parameter $f$ as a function of the conjunctive LF ($\gamma_{th}$) and the temperature of thermal electrons ($T_e$) in Figure \ref{fig: frac}. As is shown in this figure, the energy fraction is sensitively dependent on the conjunctive LF and the temperature of the thermal electrons.

\section{Models and calculations} \label{sec:cal-pol}
The structure of the GRB jet and the MF configuration of GRB prompt emission are both uncertain. 
A uniform top-hat jet is one of the candidate models. The small-scale random field and the large-scale ordered field are two competitive MF models in GRB prompt emission. The small-scale random field is usually produced by the Weibel instability
\citep{Gruzinov+Waxman+1999, Medvedev+Loeb+1999} or kinetic turbulence \citep{Mao+Wang+2011, Mao+Wang+2013} , while the large-scale ordered MF is generated from the central objects of GRB \citep{Spruit+2001}. According to the recent PIC simulations, the hybrid distribution of relativistic thermal and nonthermal electrons can be produced in a magnetic energy-dominated model \citep{Guo+etal+2014, Guo+etal+2015, Sironi+Spitkovsky+2014}. This model usually corresponds to the large-scale ordered MF structure. Hence, we consider a uniform top-hat jet with a large-scale ordered MF model in this paper. The large-scale order MF contains radial and toroidal components. The MF strength $B'$ in the jet comoving frame decreases with the jet radius as a power-law form \citep{Spruit+2001}. The radial component decreases as $B' \propto R^{-2}$, while the toroidal component decreases as $B' \propto R^{-1}$. At a large radius of the GRB emission region, the MF should be dominated by the toroidal component. The MF strength can be written as \citep{Uhm+Zhang+2014, Spruit+2001}
\begin{equation}
B'=B'_0 (\frac{R}{R_0})^{-1},
\end{equation}
where $R_0$ is the initial radius, $B'_0$ is the initial MF strength, and $R=R_0+\beta c\Gamma t'$ is the jet radius. $\Gamma$ and $\beta$ are the bulk LF and the dimensionless velocity of the jet, respectively. $t'= t_{\rm{obs}}\delta/(1+z)$ is the jet comoving time, and $t_{\rm{obs}}$ is the observed time. $\delta$ is the Doppler factor. $z$ is the redshift. For the toroidal MF model, the energy-resolved linear PD of GRB prompt emission can be calculated as \citep{Cheng+etal+2020}
\begin{equation} \label{pi-nu}
\begin{aligned}
\Pi (\nu)&=\frac{Q(\nu)}{I(\nu)}= \int_{0}^{(1+q)^{2}y_{j}} g(y) dy \int_{-\Delta \phi(y)}^{\Delta \phi(y)} d\phi  \int_{\gamma_{\rm{min}}}^{\gamma_{\rm{max}}} G(x)N_{e}(\gamma) \\
&\times B'(t') \sin \alpha' \cos (2\chi ) d\gamma  [ \int_{0}^{(1+q)^{2}y_{j}} g(y) dy\\
&\times \int_{-\Delta \phi(y)}^{\Delta \phi(y)} d\phi  \int_{\gamma_{\rm{min}}}^{\gamma_{\rm{max}}} F(x) N_{e}(\gamma) B'(t')  \sin \alpha'  d\gamma]^{-1}.
\end{aligned}
\end{equation}
where $Q(\nu)$ and $I(\nu)$ are the stokes parameters. Note that $U(\nu)=0$. According to equation \ref{pi-nu}, we can further obtain the instantaneous and time-averaged PDs in a given observed 
frequency range [$ \nu_{1}, \nu_{2} $] by using the following formula  \citep{Cheng+etal+2020, Cheng+etal+2024}
\begin{equation}\label{pi}
\begin{aligned}
\Pi &=\frac{Q}{I}=\int_{\nu_1}^{\nu_2}d\nu \int_{0}^{(1+q)^{2}y_{j}} g(y) dy \int_{-\Delta \phi(y)}^{\Delta \phi(y)} d\phi  \int_{\gamma_{\rm{min}}}^{\gamma_{\rm{max}}} G(x)N_{e}(\gamma) \\
&\times B'(t') \sin \alpha' \cos (2\chi ) d\gamma  [\int_{\nu_1}^{\nu_2}d\nu \int_{0}^{(1+q)^{2}y_{j}} g(y) dy\\
&\times \int_{-\Delta \phi(y)}^{\Delta \phi(y)} d\phi  \int_{\gamma_{\rm{min}}}^{\gamma_{\rm{max}}} F(x) N_{e}(\gamma) B'(t')  \sin \alpha'  d\gamma]^{-1}.
\end{aligned}
\end{equation}
Some variables are defined here: $y \equiv (\Gamma \theta)^{2}$,  $y_{j} \equiv (\Gamma
\theta_{j})^{2}$, and $q \equiv \theta_{v} / \theta_{j}$, where $\theta_{j}$ is the jet opening angle, and $\theta_{v}$ is the viewing angle. $\alpha'$ is the angle between the MF and the electron velocity. $\gamma_{\rm{max}}$ is the maximum LF of the electrons. For the time-resolve PD, $g(y)=(1+y)^{-3}$, while for the time-averaged PD, $g(y)=(1+y)^{-2}$ \citep{Nakar+etal+2003}. By taking the electron distribution (equation \ref {ele-dis}) into equation \ref{pi}, we can calculate the instantaneous and time-averaged PD of GRB prompt emission. $F(x)$ and $G(x)$ are written as \citep{Rybicki+Lightman+1979}
\begin{equation}
\begin{cases}
F(x)=x\int_{x}^{\infty}K_{5/3}(\xi)d\xi \\
G(x)=xK_{2/3}(x),
\end{cases}
\end{equation}
where $K_{5/3}(\xi)$ and $K_{2/3}(x)$  are Bessel functions, $x=\nu'/\nu'_{c}$, and $\nu'_c=\frac{3q_{e}B'\sin\alpha'}{4\pi m_e c}\gamma_{e}^{'2}$. Note that $\nu'$ is the frequency in the jet comoving frame.
Other variables are as follows \citep{Granot+2003, Granot+konigl+2003, Granot+Taylor+2005, Toma+etal+2009}:
\begin{equation}
\sin \alpha' = \left[ \left(\frac{1-y}{1+y}\right)^{2} + \frac{4y}{(1+y)^{2}} \frac{(s-\cos \phi)^{2}}{(1+s^{2}-2s \cos \phi)} \right]^{1/2},
\end{equation}
\begin{equation}
\chi = \phi + \rm{arctan} \left( \frac{(1-y)}{(1+y)} \frac{\sin \phi}{(s- \cos \phi)}\right),
\end{equation}
\begin{equation}
\Delta \phi(y) =
\begin{cases}
0,    \qquad \qquad \qquad \qquad \rm{for} \; q>1  \; \rm{and}  \;  y<(1-q)^2 y_{j}     \\
\pi,  \qquad \qquad \qquad \qquad \rm{for} \; q<1  \; \rm{and}  \;  y<(1-q)^2 y_{j}   \\
\cos^{-1} \left[\frac{(q^{2}-1)y_{j}+y}{2q \sqrt{y_{j}y}}\right]   \qquad \qquad \qquad \qquad \quad \rm{otherwise}.
\end{cases}
\end{equation}
where $s=\theta/\theta_v$.

\subsection{Energy-resolved polarization } \label{subsec:energy-res}
\begin{figure}[ht!] 
\plotone{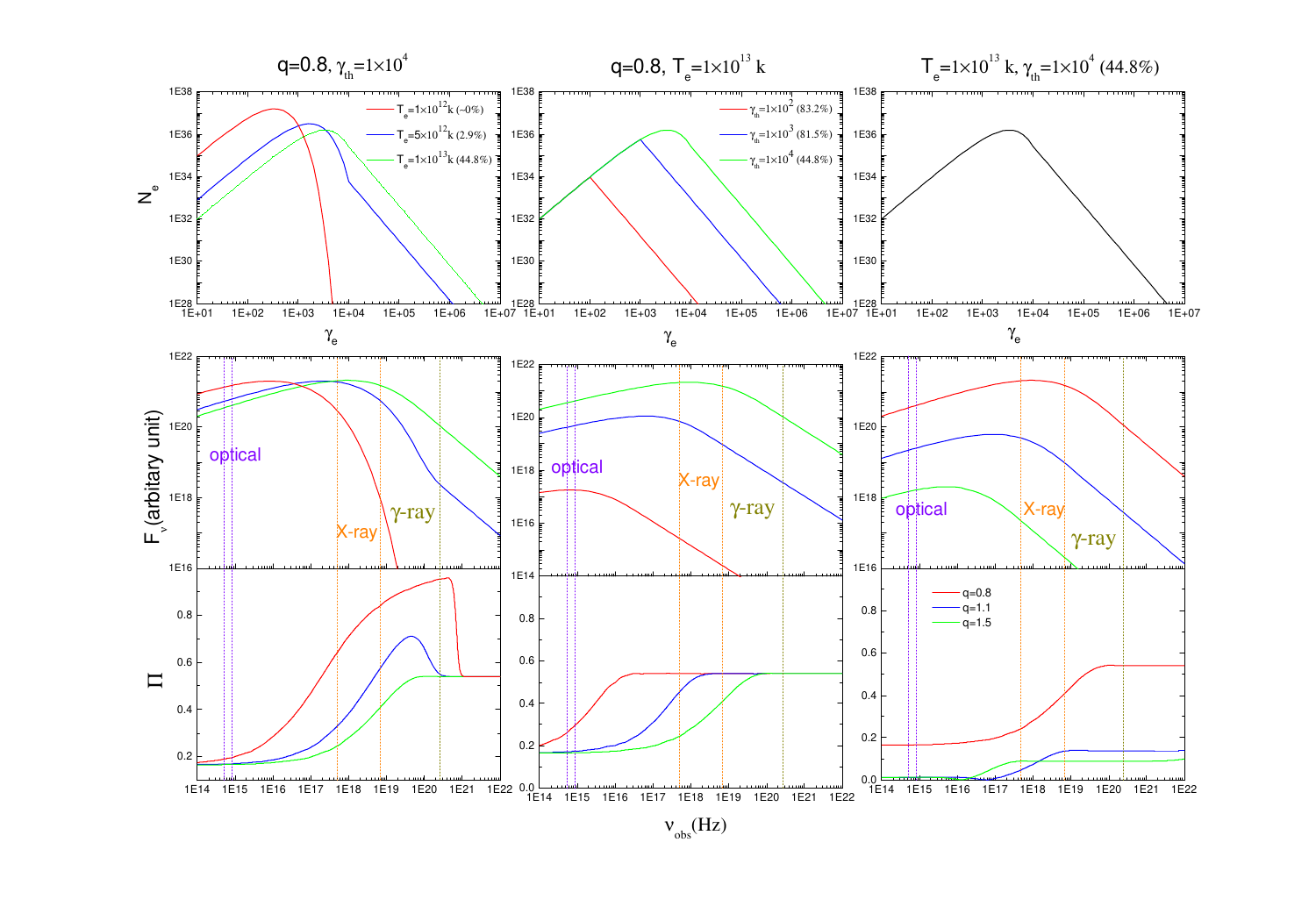}
\caption{
Polarization spectra
(bottom panels), flux spectra (middle row panels), and 
electron distributions (top panels) for
different temperatures $T_e$ (left panels), conjunctive LFs $\gamma_{th}$ (middle column panels), and normalized viewing angles $q$ (right panels). The numbers in the parentheses represent the energy fraction of nonthermal electrons. \label{fig:res-nu}}
\end{figure}

To study the influence of relativistic thermal electrons on the PDs in each energy band, we calculate the energy-resolved PDs (time-averaged) with the hybrid electrons. The energy-resolved PDs, the flux density, and the electron distributions are shown with different temperatures $T_e$, conjunctive LFs $\gamma_{th}$, and normalized viewing angles $q$ in Figure \ref{fig:res-nu}. In the left panels, we fix $q=0.8$ and $\gamma_{th}=1\times 10^4$. Then, we find that the energy-resolved PDs are significantly infected by the temperature of thermal electrons $T_e$. The PD plateaus of the energy-resolved polarization curves are dominated by the nonthermal electrons, and the rest of the curves are dominated by the thermal electrons. The PD spectrum displays a significant bump in the $\gamma$-ray and X-ray bands with the PDs higher than $\sim60\%$ if the thermal peak energy of electrons is much smaller than the conjunctive energy of electrons between the thermal and non-thermal distribution (see the left bottom panel). The high PD ($\gtrsim 60\%$) originates from the exponential decay part of the thermal component. One consequence of this is the lower the temperature, the higher the highest PD. This is because: the lower the temperature, the smaller the thermal peak energy of electrons.
In the middle column panels, we fix $q=0.8$ and $T_e = 1\times 10^{13}$ k.
As is shown in the panels, the smaller the conjunctive LF numbers the higher the PDs in the low-energy bands. This is because the smaller the conjunctive LFs the larger the radiation windows of nonthermal electrons. In the right panels, we fix $T_e = 1\times 10^{13}$k and $\gamma_{th}=1\times 10^{4}$. 
We find that the polarization spectra for the case of the off-beaming view (with $q=1.1$ and $1.5$) would be flattened compared to the on-beaming view ($q=0.8$) case. This can be explained as follows. In the on-beaming case, the radiation we received is mainly from the $1/\Gamma$ region in which the magnetic field is roughly aligned, thus it leads to a high net PD. Conversely, we received radiation from a wider region (so-called "high latitude radiation") of the jet in the off-beaming case, The polarization of the photons generated by different regions will be partially offset, thus producing a lower net PD compared to the on-beaming case.

\subsection{Instantaneous polarization} \label{subsec:time-resolve}
\begin{figure}[ht!]
\plotone{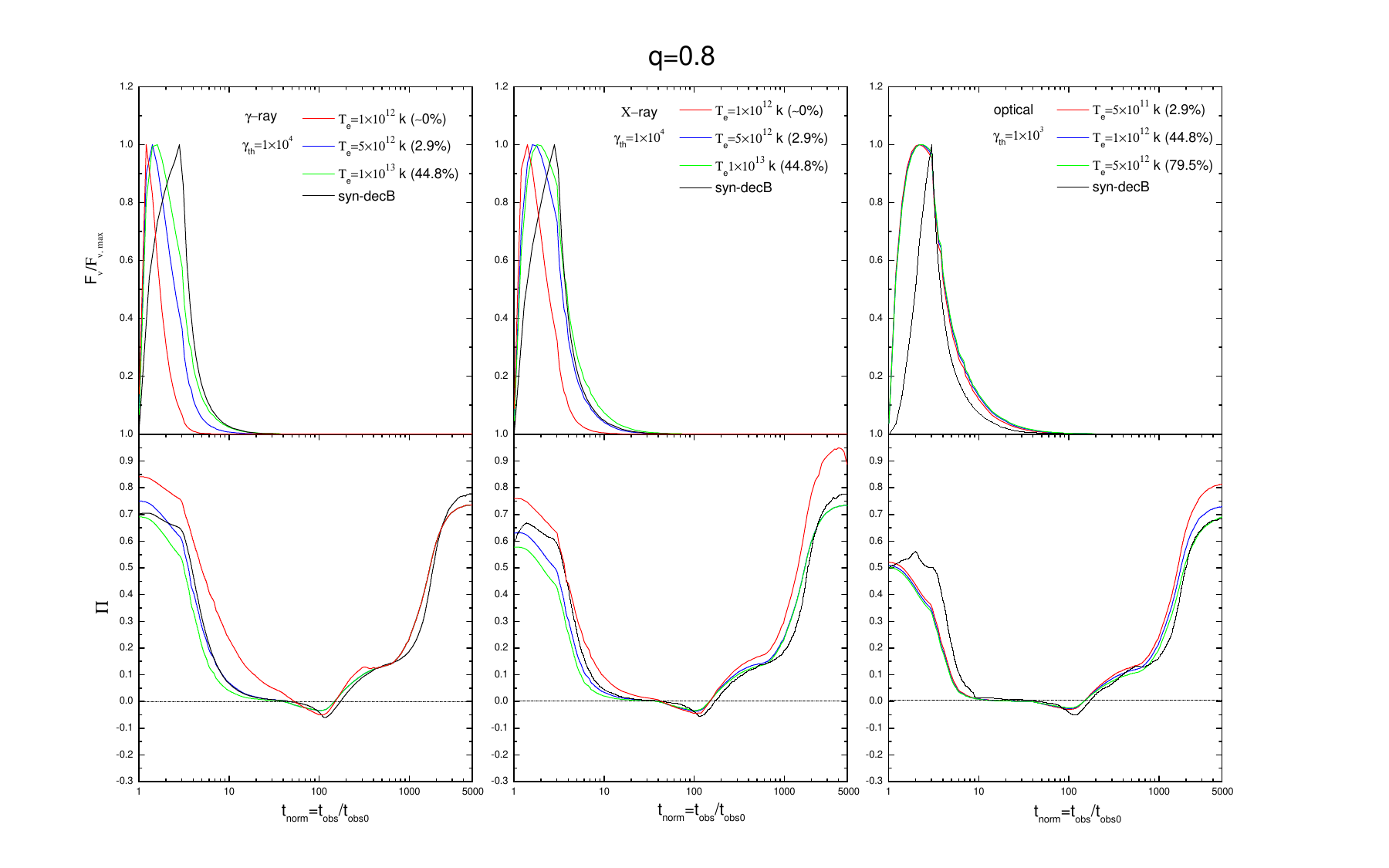}
\caption{Instantaneous PDs with different temperatures ($T_e$) in the $\gamma$-ray (30-800 keV), X-ray (2-30 keV), and optical bands ($4.3\times 10^{14} - 7.5\times 10^{14} $Hz). The upper panels display the normalized light curves (scaled to the maximum flux), while the bottom panels show the corresponding instantaneous PDs. The numbers in the parentheses represent the energy fraction of nonthermal electrons. The syn-decB represents the synchrotron radiation produced by the pure nonthermal electrons with the decaying MF model. \label{fig:res-cgth}}
\end{figure}

\begin{figure}[ht!]
\plotone{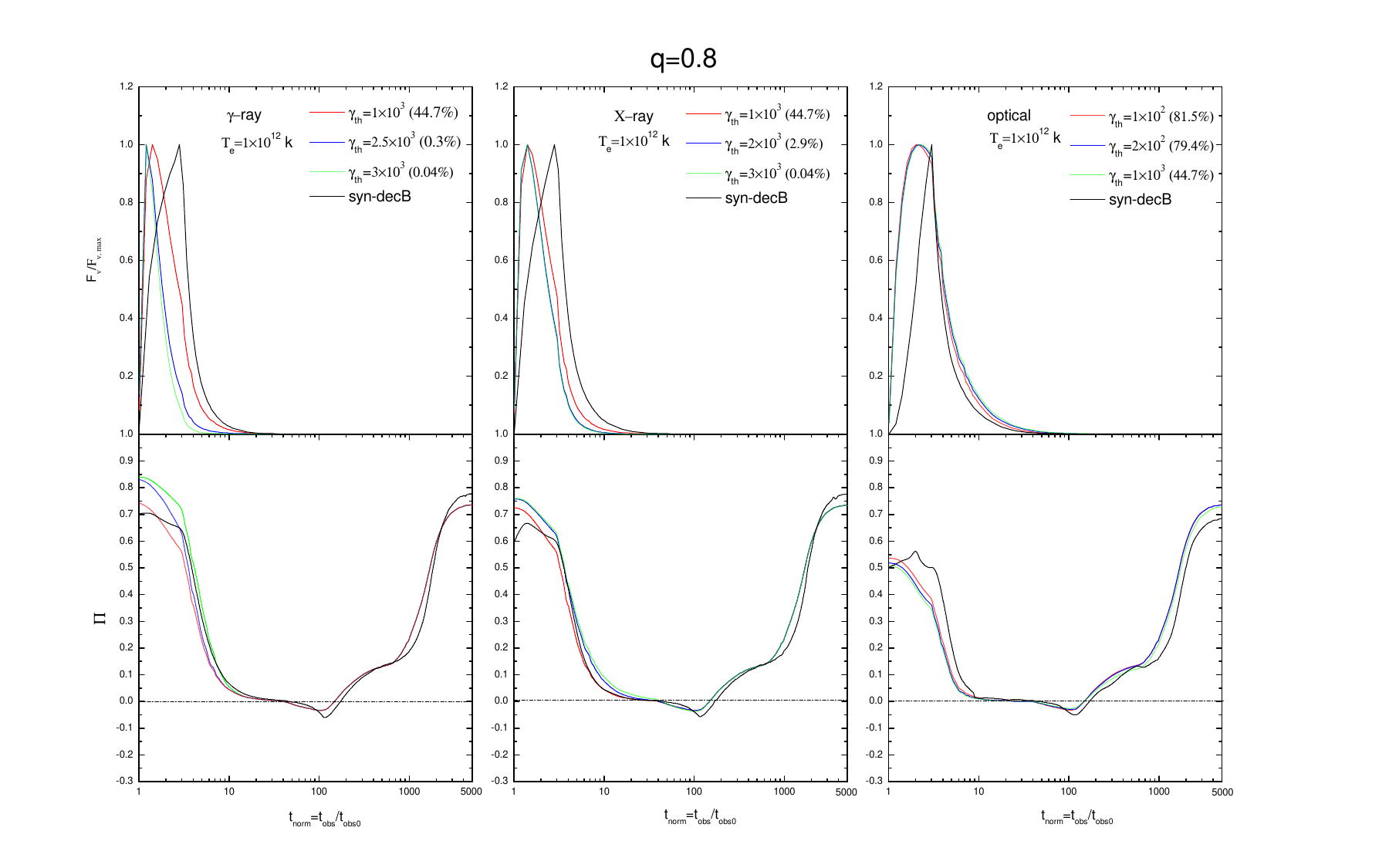}
\caption{Similar to Figure \ref{fig:res-cgth}, but with different conjunctive LFs ($\gamma_{th}$). \label{fig:res-cTe}}
\end{figure}

We calculate the instantaneous PDs in a single pulse in the $\gamma$-ray (30-800 keV), X-ray (2-30 keV), and optical bands ($4.3\times 10^{14} - 7.5\times 10^{14} $Hz). The $\gamma$-ray and X-ray bands in this paper correspond to the energy ranges of the Low-energy Polarimetry Detector (LPD, 2-30 keV) and the High-energy Polarimetry Detector (HPD, 30-800 keV) in POLAR-2, respectively. The optical band corresponds to the detected wavebands of the optical polarimeter RINGO3 \citep{Arnold+etal+2012}. Instantaneous PDs with different temperatures ($T_e$) and conjunctive LFs ($\gamma_{th}$) in the three wavebands are shown in Figure \ref{fig:res-cgth} and \ref{fig:res-cTe}. We assume a thin relativistic shell emits photons within a radius range of $R_0$ to $R_{\rm{off}}$, where $R_0$ is the radius where emission starts and $R_{\rm{off}}$ is the radius where emission turns off. The starting times of the emission are $t_{\rm{obs0}}=R_{0}(1+z)[1-\beta]/(\beta c)$ and $t_{\rm{obs0}}=R_{0}(1+z)[1-\beta \cos(\theta_{v}-\theta_{j})]/(\beta c)$ for the on-beaming and off-beaming cases, respectively. The time $t_{\rm{norm}}$ in the X-axis is normalized to the starting time $t_{\rm{obs0}}$. The turn-off times of the emission are $t_{\rm{off}}=R_{\rm{off}}(1+z)[1-\beta]/(\beta c)$ and $t_{\rm{off}}=R_{\rm{off}}(1+z)[1-\beta \cos(\theta_{v}-\theta_{j})]/(\beta c)$ for the on-beaming and off-beaming cases, respectively. We take $R_{\rm{off}}=3R_0$ in this paper, thus we have $t_{\rm{off}}=3 t_{\rm{obs0}}$. Moreover, the synchrotron polarization produced by purely nonthermal electrons with the decaying MF model \citep{Cheng+etal+2020} is added to the figures for comparison, and we mark the model as "syn-decB". As is shown in Figure \ref{fig:res-cgth}, the PDs with different temperatures display significant differences before the peak time in $\gamma$-ray and X-ray bands, while the difference in the optical band is small. However, the PD evolution for the electron temperature and that for the syn-decB model are similar. Therefore, the evolution of polarization should be dominated by geometric effects and MF structures, rather than radiation mechanisms. Moreover, the lower the temperatures (or the lower the nonthermal energy fraction) the higher the initial PDs for the three wavebands. 
The PDs obtained from the hybrid electron energy distribution could be higher than those obtained from the syn-decB model only for the cases of the electron temperature lower than $\sim 10^{12}$k. In Figure \ref{fig:res-cTe}, the temperature of the thermal electrons is fixed as $T_e =1\times 10^{12}$ k.
Similar to the PD evolution shown in Figure \ref{fig:res-cgth}, the PD evolution for the conjunctive LFs and that for the syn-decB model are similar. The larger the conjunctive LFs (or the lower the nonthermal energy fraction) the higher the initial PDs in the $\gamma$-ray and X-ray bands, while the cases are in contrast in the optical band.

\subsection{Time-averaged polarization}\label{subsec:general}
The instantaneous PDs have been given in section \ref{subsec:time-resolve}. In order to further explore the physical characteristics of synchrotron polarization for GRBs, we calculate the time-averaged PDs in a single pulse with different parameters in different wavebands in this section.

\subsubsection{Time-averaged PDs with different viewing angles}

\begin{figure}[ht!]
\plotone{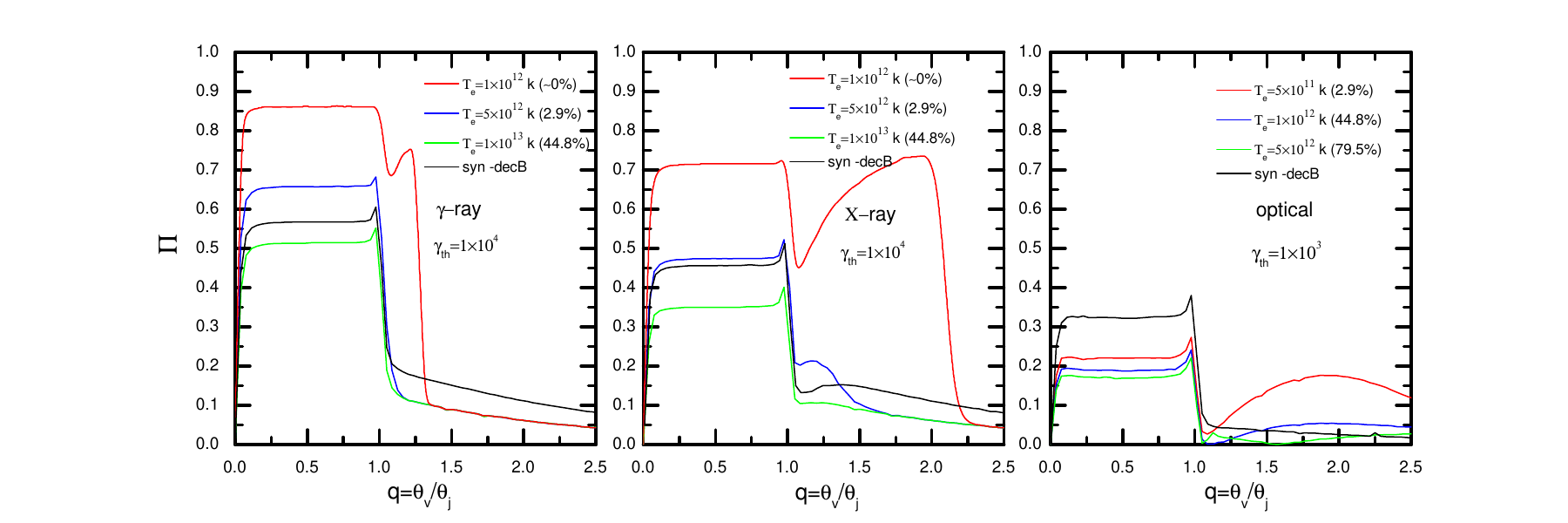}
\caption{Time-averaged PDs with different normalized viewing angles ($q$) for various $T_e$ in the $\gamma$-ray, X-ray, and optical bands. The numbers in the parentheses represent the energy fraction of nonthermal electrons. The syn-decB represents the synchrotron radiation produced by the pure nonthermal electrons with the decaying MF model. These time-averaged results correspond to the instantaneous results in Figure \ref{fig:res-cgth}. \label{fig:int-cgth}}
\end{figure}

\begin{figure}[ht!]
\plotone{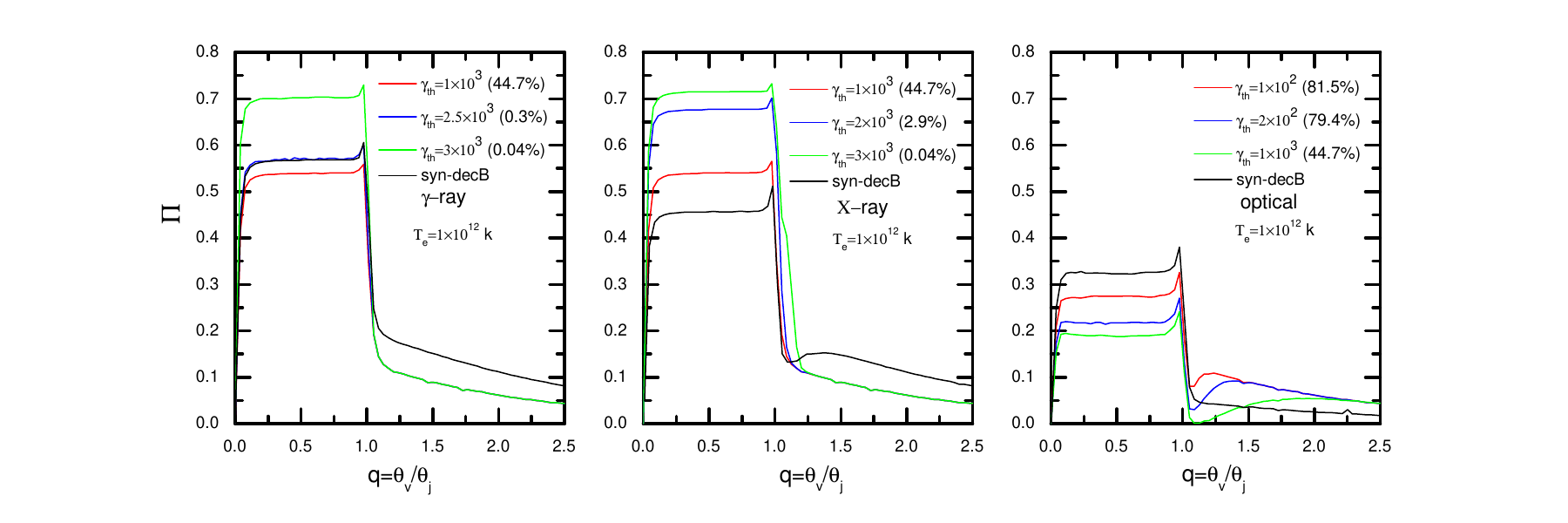}
\caption{Similar to Figure \ref{fig:int-cgth}, but with various $\gamma_{th}$. These time-averaged results correspond to the instantaneous results in Figure \ref{fig:res-cTe}. \label{fig:int-cTe}}
\end{figure}

We calculate the time-averaged PDs in the $\gamma$-ray, X-ray, and optical bands. Time-averaged PDs with the different normalized viewing angles ($q$) in various $T_e$ and $\gamma_{th}$ in the three wavebands are shown in Figure \ref{fig:int-cgth} and \ref{fig:int-cTe}. Note that we take the same physical parameters to obtain  
Figures \ref{fig:int-cgth} and \ref{fig:res-cgth}, and the same physical parameters to obtain Figures \ref{fig:int-cTe} and \ref{fig:res-cTe}. In Figure \ref{fig:int-cgth}, we fix the values of conjunctive LF and change the value of parameter $T_e$. 
The profiles of the time-averaged PDs for different electron temperatures and the profiles for the syn-decB model are similar. The PDs for the on-beaming ($q\lesssim 1$) case are much higher than those for the off-beaming ($q>1$) case in the three wavebands.
Moreover, the lower the temperatures (or the nonthermal energy fraction) the higher the time-averaged PDs for the three wavebands. For the same parameters, the PDs in the $\gamma$-ray, X-ray, and optical bands decrease successively. Compared to the syn-decB model, the PDs could be higher than those of syn-decB model only for the electron temperatures lower than $\sim 5\times 10^{12}$k in the $\gamma$-ray and X-ray bands, while the PDs in the optical band for the syn-decB model is higher than the PDs obtained from the hybrid
electrons with the temperatures $T_e \gtrsim 5\times 10^{11}$k. In Figure \ref{fig:int-cTe}, we fix the value of the electron temperature and change the value of $\gamma_{th}$. We find that the larger the $\gamma_{th}$ (or the lower the nonthermal energy fraction) the higher the time-averaged PDs in the $\gamma$-ray and X-ray bands, while the cases are in contrast in the optical band. These time-averaged results in Figures \ref{fig:int-cgth} and Figures \ref{fig:int-cTe} are consistent with the instantaneous results in Figure \ref{fig:res-cgth} and Figure \ref{fig:res-cTe}, respectively.

\subsubsection{Time-averaged PDs with different conjunctive LFs and temperatures} \label{subsec:general}

\begin{figure}[ht!]
\begin{center}
		\includegraphics[angle=0,width=1.0\textwidth, height=0.28\textheight]{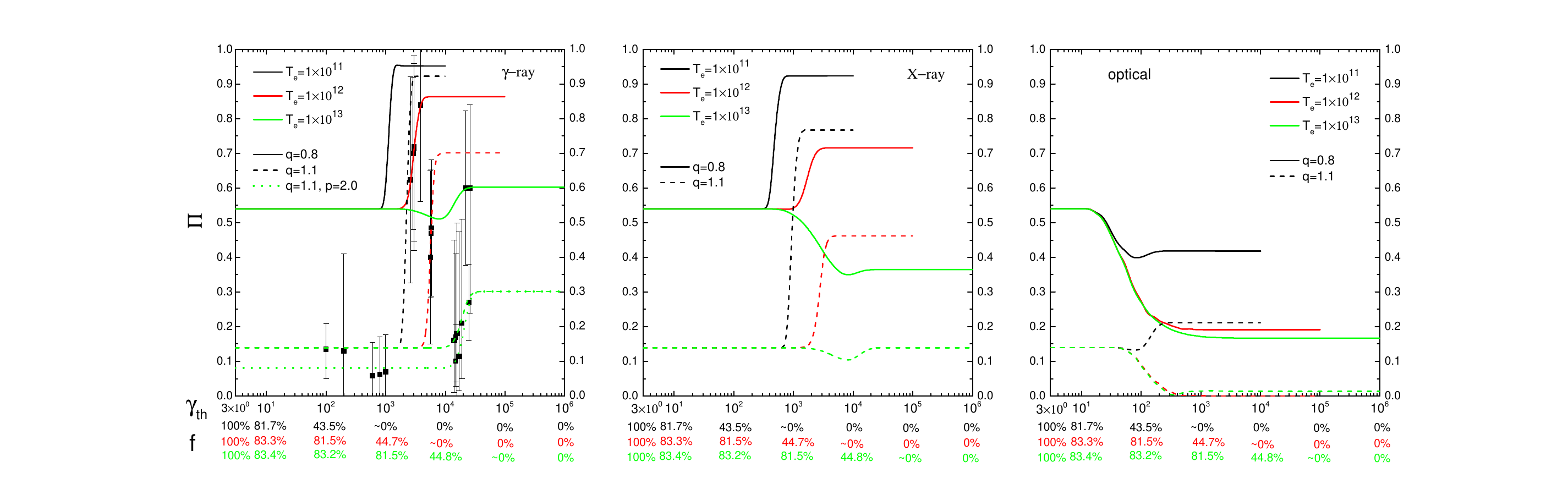}
		\end{center}
\caption{Time-averaged PDs with different $\gamma_{th}$ in various $T_e$ in the $\gamma$-ray, X-ray, and optical bands. For comparison, the observed polarization data of the $\gamma$-ray band is presented in the left panel. $f$ is the energy fraction of the nonthermal electrons. The black, red, and green numbers correspond to the temperatures of $1\times 10^{11}$ k, $1\times 10^{12}$ k, and $1\times 10^{13}$ k, respectively. The solid lines represent $q=0.8$ and the dash lines represent $q=1.1$. The dotted green line in the left panel represents $p=2.0$ and $\rm{T_e =1\times 10^{13} k}$ for $q=1.1$. Note that all curves except the dotted green line are calculated by taking p=2.8. \label{fig:pol-cTe-chgth}}
\end{figure}

In order to study the relations of PDs and the parameters of conjunctive LFs $\gamma_{th}$ and temperatures $T_e$, we calculate the time-averaged PDs with different $\gamma_{th}$ in various $T_e$ in the $\gamma$-ray, X-ray, and optical bands. The results are shown in Figure \ref{fig:pol-cTe-chgth}. For comparison, we present the observed polarization data in the $\gamma$-ray band in this Figure. The PD and spectral data are listed in Table \ref{tab: data}. According to our results, we can explain the polarization data in general. The PDs vary greatly at different temperatures in the three wavebands. In the $\gamma$-ray band, the PDs higher than $60\%$ can be produced by the electrons with the nonthermal energy fraction of $f \lesssim 10\%$, the temperatures of thermal electrons $T_e \sim [10^{11}-10^{13}]$, and the conjunctive LFs $\gamma_{th} \sim [10^{3}-10^{4}]$, respectively. Moreover, the high PDs ($> 60\%$) in the X-ray band also require a low nonthermal energy fraction ($\lesssim 10\%$), which is similar to the $\gamma$-ray band, while the optical band is hard to reach such high PDs.
The corresponding energy fraction for each PD range in the $\gamma$-ray band is as follows. 
The extremely high PDs ($\gtrsim 80\%$) can be explained by the hybrid electrons with a tiny nonthermal energy fraction ($\lesssim 0.1\%$). Both the high PDs ($60\% < \Pi < 80\%$) and the medium PDs ($20\% \lesssim \Pi \lesssim 40\%$) can be produced by the hybrid electrons with a low nonthermal energy fraction ($\lesssim 10\%$). To better explain the PDs lower than $\sim 10\%$ observed by POLAR, we calculate the PD with a harder nonthermal electron spectrum ($p=2.0$) for $q=1.1$ in the $\gamma$-ray band. By combining the results in the three wavebands, we find that the PDs higher than $\sim 10\%$ could be explained by both the on-beaming ($q\lesssim 1$) and off-beaming ($q>1$) cases with the hybrid distribution of electrons, while the PDs lower than $\sim 10\%$ could be only interpreted by the off-beaming case.

\begin{figure}[ht!]
\plotone{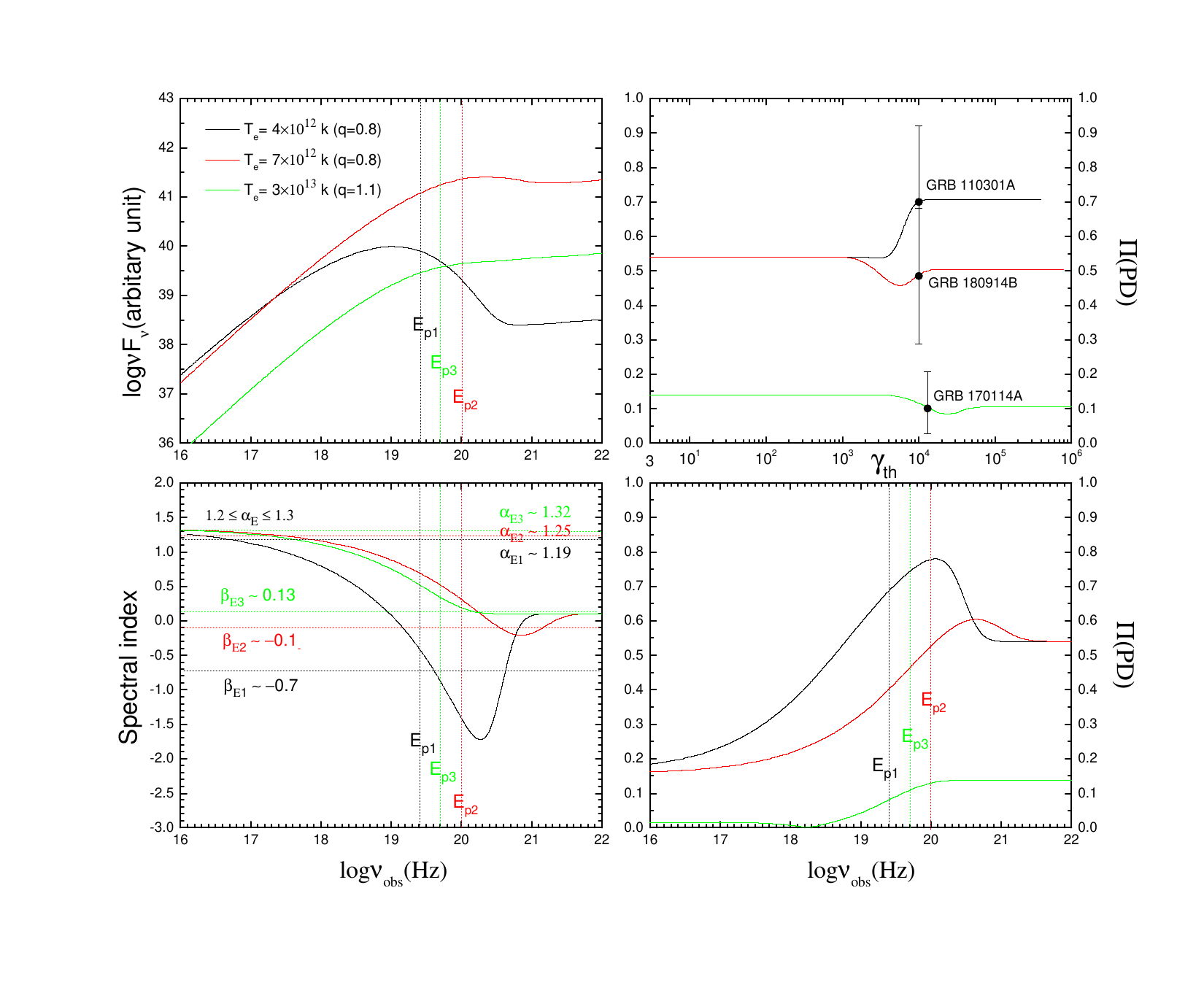}
\caption{The PD and spectral data of GRBs compared with our numerical results. The black, red, and green lines are compared with GRB 110301A, 180914B, and 170114A, respectively. For the detailed model parameters of these lines see Table \ref{tab: parameters}. The spectral parameters ($\alpha_E$, $\beta_E$, and $E_p$) with subscripts of 1, 2, and 3 correspond to GRB 110301A, 180914B, and 170114A, respectively. The observed peak energy $E_{\rm{p1}}=106.80_{+1.85}^{-1.75}$ keV, $E_{\rm{p2}}=453_{-24}^{+26}$ keV, and $E_{\rm{p3}}=211_{-25}^{+31}$ keV, respectively. The top left panel shows the flux density spectra and the bottom left panel displays the corresponding spectral index. The top right panel shows the PD with different conjunctive LFs. The bottom right panel displays the polarization spectrums. \label{fig:spec-pd}}
\end{figure}

In order to better constrain the model, in addition to the PD data, we incorporate spectral data to impose stricter restrictions on the model. The observed PD and spectral data of GRBs are compared with our numerical results in Figure \ref{fig:spec-pd}. Note that $\alpha$ and $\beta$ are the low-energy and high-energy photon indices ($N_{\nu} \propto \nu^{\alpha}$ or $\propto \nu^{\beta}$), while $\alpha_E$ and $\beta_E$ are the low-energy and high-energy spectral indices of energy spectrum ($\nu F_{\nu} \propto \nu^{\alpha_{E}}$ or $\propto \nu^{\beta_E}$). They meet the relationships of $\alpha_{E}=\alpha+2$ and $\beta_{E}=\beta+2$. $E_{p}$ is the observed peak energy of the energy spectrum. As shown in the bottom left panel, the low-energy spectral index $\alpha_{E}$ is determined by the distribution of relativistic thermal electrons. It is satisfied with $1.2\lesssim \alpha_{E} \lesssim 1.3$ ($-0.8 \lesssim \alpha \lesssim -0.7$) for different model parameters in our model. Therefore, we can select three GRBs that conform to this feature based on the spectral data given in Table \ref{tab: data}. They are GRB 110301A, 180914B, and 170114A, respectively. Their low-energy spectral indices are $\sim 1.19$, $\sim 1.25$, and $\sim 1.32$, respectively. We calculate the time-averaged PDs (varying with conjunctive LFs) and compare them with the PD data of the three GRBs (see the top right panel). By adjusting the model parameters, we obtained the results roughly consistent with the data. The detailed model parameters see Table \ref{tab: parameters}. The comparison of the spectral indices is shown in the bottom left panel and the comparison of the peak energy is in the top left panel. Moreover, the energy-resolved PDs are shown in the bottom right panel. As our results show, under the model parameters we selected, the theoretical PDs are in good agreement with the observed. Meanwhile, the spectral properties are roughly consistent with the observed. Hence, the hybrid electrons model can explain the polarization and spectral properties of some GRBs. Considering that the solely non-thermal electrons model is difficult to produce a PD higher than $60\%$, GRB 110301A may be potential evidence for the existence of relativistic thermal electrons.

\section{Summary and discussion} \label{sec: conclusion}
We calculate the synchrotron polarization of the mixture distribution of relativistic thermal and nonthermal electrons with different temperatures ($T_e$) and conjunctive LFs ($\gamma_{th}$) in the $\gamma$-ray, X-ray, and optical bands. Our main results are summarized as follows.

\begin{enumerate}
\item 
The time-averaged synchrotron polarization of the hybrid distribution of relativistic thermal and nonthermal electrons can be much higher than that of single non-thermal electrons, depending on the temperatures of the thermal electrons and the conjunctive LFs.

\item In the three wavebands, the time-averaged synchrotron PD with the hybrid distribution of electrons is generally higher than $\sim 10\%$ for both the on-beaming ($q\lesssim 1$) and the slightly off-beaming ($1<q\lesssim 1.1$) cases. In contrast, the small PD below $\sim 10\%$ can be produced for slightly off-beaming ($1<q\lesssim 1.1$) with a hard nonthermal electron spectrum or pronouncedly off-beaming cases ($q>1.2$). For the same parameters, the PD is lower in the X-ray band than in the $\gamma$-ray band, and significantly lower in the optical band than in both the X-ray and $\gamma$-ray bands.

\item The time-averaged PD spectrum displays a significant bump in the $\gamma$-ray and X-ray bands with the PDs higher than $\sim60\%$ if the thermal peak energy of electrons is much smaller than the conjunctive energy of electrons between the thermal and non-thermal distribution. 

\item The hybrid electrons model can roughly explain the PD and spectral properties of GRB 110301A, 180914B, and 170114A. Considering that the high synchrotron PD ($\gtrsim 60\%$) is hard to be produced by the purely non-thermal electrons model, GRB 110301A may be potential evidence for the existence of relativistic thermal electrons.
\end{enumerate}

Our results show that the hybrid electrons can result in both high PDs and low PDs in GRB prompt emission.
The bump of the polarization spectrum in the X-ray or $\gamma$-ray bands could be higher than $\sim 60\%$. This high PD arises from the fact that much larger conjunctive LFs $\gamma_{th}$ than the thermal peak LF will lead to the emergence of the exponential part of the thermal component in the electron distribution, which corresponds to a very steep power-law index of electron distribution and thus generates a high PD. GRB 110301A  was observed with a high PD ($70_{-22}^{+22} \%$) and a steep high-energy photon index ($\beta \sim -2.7$). Thus it can be roughly explained by the hybrid electrons model. These characteristics will be further tested by high-precision polarization measurements from future instruments, such as POLAR-2 and LEAP. Moreover, We can extract from Figure \ref{fig:res-nu} that for a given conjunctive LF $\gamma_{\rm{th}}$, the lower the peak energy $E_{\rm{peak}}$ of the flux density spectra, the higher the PD in the $\gamma$-ray and X-ray bands. The correlation of $E_{\rm{peak}}$-PD will help us to further test the model. Note that these results are obtained from a uniform top-hat jet with a large-scale toroidal MF model. The polarization should be closely related to the MF structure and jet structure. In this study, we found the PD in the large-scale toroidal field with a hybrid electron distribution could be much higher than that with a pure nonthermal electron distribution. Thus we expect that the PD in the small-scale random field with the hybrid electron distribution can be also higher than that with the pure nonthermal electron distribution. Moreover, the polarization of the hybrid electrons in a structured jet should be different from the uniform top-hat jet model due to the energy structure within the jet, and the difference depends on the angular energy distribution of the structured jet. The steeper the energy distributed with the angles of the structured jets, the greater the difference in PD between the structured and uniform jets. The synchrotron polarization of the hybrid electrons in the structure jet and the small-scale random field will be investigated in further work.

Since the PDs lower than $\sim 60\%$ can be produced by various models, constraining the models in a single energy band is difficult. In the future, GRB polarization measurements in multi-wavelength would be better to constrain the temperature of the thermal electrons and the conjunctive LF in GRBs. 
Simultaneous detection of X-ray and $\gamma$-ray emissions by the polarimeter POLAR-2 \citep{Kole+etal+2024}, along with optical polarization observations from polarimeters such as MOPTOP \citep{2020MNRAS.494.4676S}, could provide new insight into particle acceleration mechanisms and the magnetic field structure in the GRB emission region.

It is important to note that our polarization results can apply to GRB X-ray flares. X-ray flares exhibit some observational features that resemble the prompt emission of GRBs (e.g., \citealt{Chincarini+etal+2010, Liu+Mao+2019}), suggesting an internal origin similar to GRB prompt emission \citep{2006ApJ...642..354Z}. \cite{Geng+etal+2018} considered synchrotron radiation from the purely non-thermal electrons and modeled the polarization evolution of some observed GRB X-ray flares. Their results show that the instantaneous PD slowly decreases from a maximum value in the rising phase of the X-ray flare, and then decreases rapidly, followed by a rapid rise, which is similar to our results. However, the initial PD in our mixed electron energy distribution could be much higher than that in their single non-thermal distributions, and the detailed instantaneous evolution of the PD in the two cases is different. The time-resolved polarization detection from the Low-energy Polarimetry Detector (LPD) in POLAR-2 will possibly distinguish between these two cases.

\begin{acknowledgments}
This work is supported by the Guangxi Natural Science Foundation (GUIKEAD22035945), the Guangxi Natural Science Foundation (2024GXNSFBA010350), the Scientific Research Project of Guangxi Minzu University (2021KJQD03), the National Key R\&D Program of China (2023YFE0101200), and the National Natural Science Foundation of China (Nos. U1831135, 12393813, 12393811). J.M is supported by the Yunnan Revitalization Talent Support Program (YunLing Scholar Project).
\end{acknowledgments}

\begin{deluxetable}{ccccccccc}[h] \label{tab: data}
	\tablecolumns{9}
	\setlength{\tabcolsep}{9pt}
	\tablewidth{0pc}
	\tablecaption{The measured PD ($\gamma$-ray band ) and spectral data of GRBs and the PD data constrain to our model parameters.}
	\tabletypesize{\scriptsize}
	\tablehead{
		\colhead{GRB}&
		\colhead{PD(\%)}&
		\colhead{$\alpha/\beta$}&
		\colhead{$E_{\rm{p}}$ (keV)}&
		\colhead{Instrument (PD)}&
		\colhead{$f$} &
		\colhead{$T_{e}$(K)}&
		\colhead{$\gamma_{\rm{th}}$} 
	}
	\startdata
	110721A & $84_{-28}^{+16}$ &$-0.94_{-0.02}^{+0.02} / -1.77_{-0.02}^{+0.02}$ & $372.50_{-23.60}^{+26.50}$  & IKAROS-GAP  &$ [0\% - 0.1\%] $&$[10^{11}-10^{12}]$ &$[10^{3}-10^{4}]$ \\
	180720B & $72_{-30}^{+24}$ &$-1.10_{-0.01}^{+0.01} / -2.24_{-0.03}^{+0.03}$ &$747_{-25}^{+25}$  & $(*)$ & $ [0\% - 10\%]$&$ [10^{12}-10^{13}]$ &$ [10^{3}-10^{4}]$ \\
	180103A & $71.4_{-26.8}^{+26.8}$ &$-1.31_{-0.06}^{+0.06} / -2.24_{-0.13}^{+0.90}$ &$273_{-23}^{+26}$  & AstroSat-CZTI & $ [0\% - 10\%]$&$ [10^{12}-10^{13}]$ &$ [10^{3}-10^{4}]$ \\
	110301A & $70_{-22}^{+22}$ &$-0.81_{-0.02}^{+0.02} / -2.70_{-0.05}^{+0.04}$ &$106.80_{+1.85}^{-1.75}$ & IKAROS-GAP& $ [0\% - 10\% ]$&$[10^{12}-10^{13}]$ &$[10^{3}-10^{4}]$  \\
	180120A & $62.4_{-29.8}^{+29.8}$ &$-1.01_{-0.014}^{+0.014} / -2.4_{-0.09}^{+0.09}$ & $140.91_{-3}^{+3}$ & AstroSat-CZTI & $[0\% - 10\%]$&$[10^{12}-10^{13}]$ &$[10^{3}-10^{5}]$\\
	180427A & $60.0_{-22.3}^{+22.3}$ &$-0.29_{-0.077}^{+0.077} / -2.80_{-0.16}^{+0.16}$ &$147_{-2}^{+2}$  & AstroSat-CZTI &$[0\% - 100\%]$&$[10^{11}- 10^{13}]$ &$ [\gamma_{\rm{min}}-\gamma_{\rm{max}}]$ \\
	170101B & $60_{-36}^{+24}$ &$-0.59_{-0.05}^{+0.05} / -3.28_{-0.49}^{+0.46}$ & $232_{-10}^{+10}$ & POLAR & $[0\% - 100\%]$&$[10^{11}- 10^{13}]$ &$ [\gamma_{\rm{min}}-\gamma_{\rm{max}}]$  \\
	180914B & $48.5_{-19.7}^{+19.7}$ &$-0.75_{-0.04}^{+0.04} / -2.1_{-0.70}^{+0.08}$ &$453_{-24}^{+26}$  & AstroSat-CZTI & $ [0\% - 100\%]$&$[10^{11}- 10^{13}]$ &$[\gamma_{\rm{min}}-\gamma_{\rm{max}}]$  \\
	190530A & $46.9_{-18.5}^{+18.5}$ &$-0.99_{-0.002}^{+0.022} / -3.50_{-0.25}^{+0.25}$ &$888_{-8}^{+8}$  & AstroSat-CZTI & $ [0\% - 100\%]$ & $[10^{11}- 10^{13}]$ &$ [\gamma_{\rm{min}}-\gamma_{\rm{max}}]$ \\
	170305A & $40_{-25}^{+25}$ &$-0.35_{-0.09}^{+0.09} / -3.20_{-0.49}^{+0.41}$ &$253_{-16}^{+17}$  & POLAR & $ [0\% - 10\%]$&$ [10^{11}-10^{13}]$ &$[10^{3}-\gamma_{\rm{max}}]$ \\
	100826A& $27_{-11}^{+11}$ &$-1.31_{-0.05}^{+0.06} / -2.1_{-0.2}^{+0.1}$ &$606_{-109}^{+134}$ & IKAROS-GAP & $ [0\% - 10\%]$ & $[10^{11}-10^{13}]$ &$ [10^{3}-\gamma_{\rm{max}}]$ \\
	161217C & $21_{-16}^{+30}$ &$-1.08_{-0.25}^{+0.43} / -2.76_{-0.61}^{+0.36}$ &$143_{-34}^{+37}$ &POLAR & $ [0\% - 10\%]$ & $  [10^{11}-10^{13}]$ &$[10^{3}-\gamma_{\rm{max}}]$  \\
	170320A & $18_{-18}^{+32}$ &$-0.24_{-0.17}^{+0.13} / -2.32_{-0.21}^{+0.16}$ &$228_{-15}^{+13}$ & POLAR  & $ [0\% - 100\%]$& $ [10^{11}-10^{13}]$ &$ [\gamma_{\rm{min}}-\gamma_{\rm{max}}]$ \\
	161229A & $17_{-13}^{+24}$ &$-0.64_{-0.03}^{+0.03} / -3.07_{-1.49}^{+0.72}$ &$339_{-14}^{+12}$ & POLAR  & $[0\% - 100\%]$ & $ [10^{11}-10^{13}]$ &$ [\gamma_{\rm{min}}-\gamma_{\rm{max}}]$\\
	161203A & $16_{-15}^{+29}$ &$0.13_{-0.25}^{+0.27} / -3.41_{-0.46}^{+0.39}$ &$344_{-12}^{+19}$ & POLAR & $ [0\% - 100\%]$ &$ [10^{11}-10^{13}]$ &$ [\gamma_{\rm{min}}-\gamma_{\rm{max}}]$  \\
         170206A & $13.5_{-8.6}^{+7.4}$ &$-0.49_{-0.03}^{+0.04} / -2.68_{-0.19}^{+0.14}$ &$344_{-12}^{+13}$  & POLAR & $ [0\% - 100\%]$ &$ [10^{11}-10^{13}]$ &$ [\gamma_{\rm{min}}-\gamma_{\rm{max}}]$   \\
         161218B & $13_{-13}^{+28}$ &$-0.51_{-0.01}^{+0.01} / -3.06_{-0.10}^{+0.10}$ &$214.80_{-2.51}^{+2.51}$ & POLAR  & $ [0\% - 100\%]$ & $ [10^{11}-10^{13}]$ &$ [\gamma_{\rm{min}}-\gamma_{\rm{max}}]$ \\
         170210A & $11.4_{-9.7}^{+35.7}$ &$-0.96_{-0.02}^{+0.02} / -2.72_{-0.49}^{+0.39}$ &$462_{-22}^{+22}$  & POLAR & $ [0\% - 100\%]$ &$ [10^{11}-10^{13}]$ &$ [\gamma_{\rm{min}}-\gamma_{\rm{max}}]$  \\
         170114A & $10.1_{-7.4}^{+10.5}$ &$-0.68_{-0.09}^{+0.09} / -1.87_{-0.05}^{+0.04}$ &$211_{-25}^{+31}$ & POLAR & $ [0\% - 100\%]$ & $ [10^{11}-10^{13}]$ &$ [\gamma_{\rm{min}}-\gamma_{\rm{max}}]$ \\
         170127C & $9.9_{-8.4}^{+19.3}$ &$-1.14_{-0.22}^{+0.21} / -3.1_{-0.6}^{+0.6} $ &$1500_{-900}^{+800}$  & POLAR  & $ [0\% - 100\%]$ &$ [10^{11}-10^{13}]$&$ [\gamma_{\rm{min}}-\gamma_{\rm{max}}]$  \\
         161218A & $7.0_{-7.0}^{+10.7}$ &$-0.54_{-0.06}^{+0.07} / -2.51_{-0.15}^{+0.14}$ &$144_{-11}^{+12}$  & POLAR & $ [0\% - 100\%]$ & $ [10^{11}-10^{13}]$&$ [\gamma_{\rm{min}}-\gamma_{\rm{max}}]$  \\
         170101A & $6.3_{-6.3}^{+10.8}$ &$-1.44_{-0.17}^{+0.13} / -2.49_{-0.23}^{+0.65} $ &$123_{-21}^{+23}$  & POLAR & $ [0\% - 100\%]$ & $ [10^{11}-10^{13}]$ &$ [\gamma_{\rm{min}}-\gamma_{\rm{max}}]$  \\
         170207A & $5.9_{-5.9}^{+9.6}$ &$-0.87_{-0.02}^{+0.02} / -3.37_{-1.26}^{+0.74}$ &$475_{-20}^{+19}$ & POLAR & $ [0\% - 100\%]$ & $ [10^{11}-10^{13}]$ &$ [\gamma_{\rm{min}}-\gamma_{\rm{max}}]$ \\
	\enddata
	\tablecomments{(*) The PD of GRB 180720B is obtained from the polarimetric analysis using the Fermi Gamma-ray Burst Monitor in \cite{Veres+etal+2024}. The spectral parameters ($\alpha, \beta$, and $E_p$) of GRB 110721A, 110301A, and 100826A are obtained from \cite{Tierney+Kienlin+2011}, \cite{Foley+2011}, and \cite{Golenetskii+etal+2011}, respectively. The spectral parameters of GRB 170127C and 170101A are obtained from \cite{Guan+Lan+2023}. The rest of the spectral and PD data of IKAROS-GAP GRBs, POLAR GRBs, and AstroSat-CZTI GRBs are collected from \cite{Yonetoku+etal+2012}, \cite{Kole+etal+2020}, and \cite{Chattopadhyay+etal+2022}, respectively. The model parameters ($f$, $T_e$, and $\gamma_{th}$) are given by comparing the PD data with the numerical results of our model in Figure \ref{fig:pol-cTe-chgth}.}
\end{deluxetable}

\begin{deluxetable}{ccccccccc}[h] \label{tab: parameters}
	\tablecolumns{9}
	\setlength{\tabcolsep}{9pt}
	\tablewidth{0pc}
	\tablecaption{The model parameters which are given by the comparison of the PD and spectral data of GRBs and our model.}
	\tabletypesize{\scriptsize}
	\tablehead{
	        \colhead{Number}&
		\colhead{GRB}&
		\colhead{$\Gamma$}&
		\colhead{$B'_{0} (\times 10^{3}$ G)}&
		\colhead{$f$} &
		\colhead{$T_{e}$(K)}&
		\colhead{$\gamma_{\rm{th}}$} &
		\colhead{$p$} &
		\colhead{$q$} 
	}
	\startdata
1& 110301A & $300$ &$4$ &$0.36 \%$  & $ 4\times 10^{12}$ & $1\times 10^{4}$ & $2.8$ &$0.8$ \\
2& 180914B & $1000$ &$6$ &$18.8\%$  & $ 7\times 10^{12}$ & $1\times 10^{4}$ & $2.8$ &$0.8$ \\
3&170114A & $300$ &$8$ &$73\%$ & $3\times 10^{13}$ & $1.3\times 10^{4}$ & $2.8$ &$1.1$ \\
	\enddata
	\tablecomments{These model parameters are given by comparing the measured PD and spectral data of GRBs with the numerical results of our model in Figure \ref{fig:spec-pd}.}
\end{deluxetable}

%





\bibliography{sample631}{}
\bibliographystyle{aasjournal}



\end{document}